\documentstyle[psfig]{kapproc}
\kluwerbib
\startauthorindex

\begin{document}
\input{epsf}

\articletitle{Copper and Zinc in Bulge-Like Stars}

\author{Luciana Pomp\'eia}
\affil{IAG - USP}
\email{pompeia@astro.iag.usp.br}

\chaptitlerunninghead{Copper and Zinc in Bulge-like Stars}

\anxx{Pomp\'eia}


\begin{abstract}

The star formation history of a given population remains imprinted in the chemical
distributions of the their individuals. The parallel study of stellar abundances and 
nucleosynthesis helps to understand the production of the elements inside stars and  
their evolution in the interestellar medium. In the present paper copper and zinc
abundances of a sample of bulge-like stars are inferred and compared to samples of the
solar vicinity. A disk-like distribution of [Zn/Fe] is found for the bulge-like stars, 
while underabundant [Cu/Fe] ratios relative to disk samples are determined.

\end{abstract}


\section{Introduction}

Element production in the Universe (except for those forged by the Big Bang nucleosynthesis),
occurs mainly in stellar interiors, during their numerous evolutionary stages. The production 
and ejection cycle of such elements in the interestellar medium of galaxies remains 
imprinted in the stellar abundance distributions, and depends basically on the nucleosynthetic
processes occurred, and on the dynamical evolution of the born region. Input assumptions for the chemical
evolution models as the star formation rate (SFR) and the initial mass function (IMF), are also constrained 
by the abundance ratios of elements produced in different stellar phases, and in stars of different
stellar masses (e.g. Chiappini et al. 2001).
 
In the present work Cu and Zn abundances of a sample of bulge-like stars are reported. This sample 
is described in our previous papers (e.g. Pomp\'eia et al. 2002, Paper I, and 2003, Paper II), 
which members are old G and K dwarf stars, with isochronal ages of 10 -11 Gyr, and probable originated near the
bulge. The metallicity range of the sample is -0.80 $\leq$ [Fe/H] $\leq$ +0.40, comprising one of the most 
metal-rich samples with derived Cu and Zn abundances.

\section{Abundance Analysis}

Stellar parameters are calculated as follows. Temperatures are derived by using H$\alpha$ line profiles; trigonometric 
surface gravities are inferred by using Hipparcos paralaxes; microturbulent velocities are calculated by requiring zero 
slope of [Fe/H] vs. EW for a list of Fe I lines; and finally iron abundances are calculated using Fe II lines. The 
final parameters are given in Table 1.    

Abundances of Zn and Cu are derived by using the spectrum synthesis code described in Cayrel et al. (1991). 
Atomic data are from Sneden et al. (1991) for the Zn I line at $\lambda$ 4722.15 $\rm \AA$, Bi\'emont \& Godefroid 
(1980) for Zn I $\lambda$ 4810.53 $\rm \AA$, and from Sneden \& Crocker (1988) with hyperfine splitting structure 
and isotopic data from Biehl (1976) for the Cu I line $\lambda$ 5782.13 $\rm \AA$. The final abundance ratios relative 
to iron are also given in Table 1.


\begin{table}[ht]
\caption[Stellar Parameters and Chemical Abundances.]
{Stellar Parameters and Chemical Abundances}
\begin{tabular*}{\hsize}{@{\extracolsep{\fill}}lcccccc}
\sphline
Star&T$_{eff}$&log g&[Fe/H]&$\xi_{t}$&[Zn/Fe]&[Cu/Fe]\cr
\hline
\sphline
HD 143016&  5575  &  4.11  &  -0.35  &  0.70  &  0.11  &  -0.35   \cr
HD 143102&  5500  &  3.85  &  0.03   &  1.05  &  0.03  &  -0.10   \cr
HD 148530&  5350  &  4.43  &  0.10   &  0.80  &  -0.10 &  -0.20    \cr
HD 149256&  5350  &  3.73  &  0.34   &  0.80  &  0.10  &  0.00    \cr
HD 152391  &  5300  &  4.45  &  -0.05  &  0.50  &  -0.03 &  -0.42   \cr
HD 326583  &  5600  &  3.81  &  -0.30  &  0.60  &  0.25  &  -0.33   \cr
HD 175617  &  5550  &  4.56  &  -0.44  &  0.60  &  0.13  &  -0.22   \cr
HD 178737  &  5575  &  3.90  &  -0.35  &  0.60  &  0.23  &  -0.20   \cr
HD 179764  &  5450  &  4.26  &  0.06   &  1.10  &  -0.09 &  -0.10   \cr
HD 181234  &  5350  &  4.25  &  0.40   &  1.10  &  0.00  &  0.00    \cr 
HD 184846  &  5600  &  4.40  &  0.06   &  0.50  &  0.10  &  -0.42   \cr
BD-176035  &  4750  &  4.36  &  0.46   &  0.70  &  0.00  &  -0.15   \cr
HD 198245  &  5650  &  4.31  &  -0.60  &  0.60  &  0.11  &  -0.22   \cr
HD 201237  &  4950  &  4.08  &  0.15   &  0.50  &  -0.11 &  -0.15   \cr
HD 211276  &  5500  &  4.05  &  -0.39  &  0.50  &  -0.09 &  -0.35   \cr
HD 211532  &  5350  &  4.46  &  -0.54  &  0.80  &  0.02  &  -0.14   \cr
HD 211706  &  5800  &  4.25  &  0.16   &  0.80  &  0.00  &  -0.40   \cr
HD 214059  &  5550  &  3.81  &  -0.38  &  0.75  &  0.31  &  -0.12   \cr
CD-40 15036 & 5350  &  4.34  &  0.00   &  0.70  &  -0.05 &     -    \cr
HD 219180  &  5400  &  4.35  &  -0.46  &  0.65  &  0.00  &  -0.30   \cr
HD 220536  &  5850  &  4.17  &  -0.11  &  0.50  &  -0.05 &  -0.32   \cr 
HD 220993  &  5600  &  4.15  &  -0.16  &  0.80  &  0.14  &  -0.20   \cr
HD 224383  &  5800  &  4.14  &  -0.06  &  1.00  &  0.00  &  -0.12   \cr
HD   4308  &  5600  &  4.31  &  -0.26  &  0.80  &  0.06  &  -0.21   \cr
HD   6734  &  5000  &  3.40  &  -0.36  &  0.75  &  0.05  &  -0.20   \cr
HD   8638  &  5500  &  4.38  &  -0.29  &  0.60  &  0.10  &  -0.17   \cr
HD   9424  &  5350  &  4.35  &  0.25   &  0.70  &  -0.04 &  -0.12   \cr
HD  10576  &  5850  &  4.00  &  -0.02  &  1.25  &  -0.27 &  -0.40   \cr
HD  10785  &  5850  &  4.16  &  -0.25  &  1.20  &  -0.03 &  -0.12   \cr
HD  11306  &  5200  &  4.09  &  -0.98  &  0.80  &  -0.06 &  -0.24   \cr
HD  11397  &  5400  &  4.34  &  -0.59  &  0.65  &  -0.03 &  -0.22   \cr
HD  14282  &  5800  &  3.91  &  -0.34  &  0.65  &  0.20  &  -0.30   \cr
HD  16623  &  5700  &  4.26  &  -0.51  &  0.60  &  0.16  &  -0.25   \cr
BD-02 603  &  5450  &  3.75  &  -0.79  &  0.70  &  -0.03 &  -0.12   \cr
HD  21543  &  5650  &  4.37  &  -0.55  &  0.50  &  0.16  &  -0.22   \cr
\sphline
\end{tabular*}
\end{table}

\section{Discussion}

The production sites of Cu and Zn are still poorly understood. As iron-peak group members, they   
are predicted to be produced in massive stars by NSE (nuclear statistic equilibrium) and by explosive 
nucleosynthesis (e.g. Woosley \& Weaver 1995). Important yields are inferred from $s$-process in massive stars, 
with a smaller contribution from AGB stars (e.g. Raiteri et al. 1992, Gallino et al. 1998). SNe Ia have also 
been claimed as stellar sources of Cu and Zn, but their contribution remains unclear (Mishenina et al. 2002).   

In Figure 1, [Cu/Fe] and [Zn/Fe] trends with metallicity for the stars of our sample are depicted. 
Stars from Sneden et al. (1991), Mishenina et al. (2002) and Reddy et al. (2003) are also given. 
As shown in this Figure, our data for [Cu/Fe] ratios are underabundant relative to the disk samples for overlapping 
metallicities. Such behavior is in contrast with the $r$-process element Eu, for which we found overabundances 
relative to iron (Paper II). This result suggests a little or no contribution from SNe II for Cu abundances.
 
[Zn/Fe] vaules, on the other hand, show an overlapping behavior relative to disk stars, with a possible trend of increase 
with decreasing metallicities. For metallicities higher than solar, the abundance ratios indicate a 
flatter behavior. 

\begin{figure}[ht]
\centerline{\psfig{file=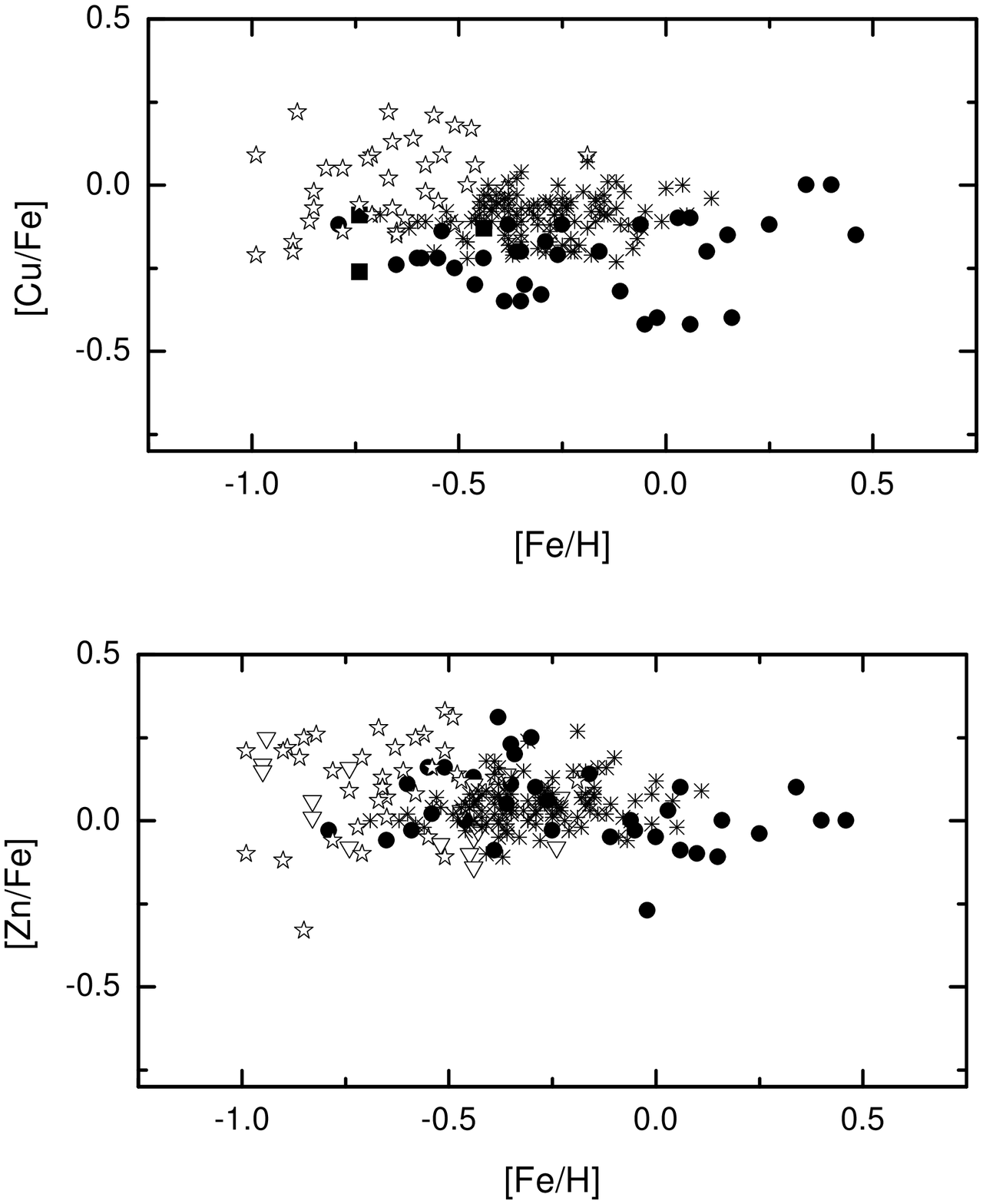,height=4in}}
\caption{[Cu/Fe] and [Zn/Fe] abundance trends with metallcity for our sample stars (solid circles), and for 
the samples of Sneden et al. (1991) (down triangles), Mishenina et al. (2002) (stars) and Reddy et al. (asterisks).}
\end{figure}


\begin{acknowledgments}
\small
I would like to thank B. Barbuy for the revision of this paper and G. Wallerstein and R. Gallino for 
very useful discussions. FAPESP pos-doc fellowship n$^{\rm o}$ 01/14594-2 and FAPESP project n$^{\rm o}$ 
1998/10138-8 are acknowledged.
\end{acknowledgments}

\end{document}